ARTICLE

# Temporal multilayer structures for designing higher-order transfer functions using time-varying metamaterials

D. Ramaccia[1,a], A. Alù[2,3], A. Toscano[1], F. Bilotti[1]

AFFILIATIONS

[1] Department of Engineering, Roma Tre University, Rome, I-00146, Italy
[2] Photonics Initiative, Advanced Science Research Center, New York, NY 10031, USA
[3] Department of Electrical Engineering, City College, City University of New York, New York, NY 10031, USA

[a] Author to whom correspondence should be addressed: davide.ramaccia@uniroma3.it

ABSTRACT

Temporal metamaterials are artificial materials whose electromagnetic properties change over time. In analogy with spatial media and metamaterials, where their properties change smoothly or abruptly over space, temporal metamaterials can exhibit a smooth variation over time, realizing a temporal non-homogeneous medium, or a stepwise transition, realizing the temporal version of dielectric slabs or multilayer structures. In this Letter, we focus our attention on temporal multilayer structures, and we propose the synthesis of higher-order transfer functions by modeling the wave propagation through a generalized temporal multilayer structure, consisting of a cascade over time of different media. The tailoring of the scattering response of temporal structure as a function of frequency is presented, deriving the corresponding scattering coefficients for a properly designed set of medium properties, *i.e.*, permittivity and permeability, and application time, in analogy with what is typically done in optical and electromagnetic spatial multilayered structures. This allows us to design novel electromagnetic and optical devices with higher-order transfer functions by exploiting the temporal dimension instead of the spatial one.

*Information of the publisher:* https://doi.org/xx.xxxxxx

Time-varying metamaterials have been attracting a large interest in the scientific community thanks to their possibility to exploit the temporal dimension for unlocking novel and unprecedented light-matter interactions. Periodic time-varying profiles of material properties have been exploited to achieve non-reciprocity [1–8], generate/control frequency harmonics [4,9], and control Doppler shift and compensation [10,11], thanks to the synthetic motion of a medium in one specific direction [2,5,6,12–15]. This functionality has been also applied to surfaces, to enable frequency translation through the modulation of transmitting [16–18] and reflecting [15,19,20] metasurfaces, or through time refraction in an ENZ material [21], and to support the analogue of Wood anomalies though surface-wave excitation in a time grating [22].

More recently, abrupt changes in time of the material permittivity and/or permeability have been also investigated as an alternative strategy to achieve instantaneous frequency conversion of the propagating waves. The sudden change of the medium properties realizes the temporal counterpart of the well-known spatial interface between two different media. In the case of spatial interfaces, the frequency is conserved across the interface, whereas the spatial distribution of the wave is not, leading to a change of wavelength, and thus wavevector, proportional to the difference of the refractive indices before and after the interface [23,24]. As a dual scenario, in the case of temporal interfaces the spatial distribution is conserved, whereas the temporal frequency is instantaneously shifted in frequency to satisfy the $\omega - k$ relationship in the new medium [25–28]. Some similarities can be also identified in the scattering response of a temporal interface with respect to a spatial one. In spatial interfaces, Fresnel equations [24] describe the relationship between the amplitudes of the reflected and refracted electromagnetic waves with respect to the incident one. Dual relations have been derived in [25,26] for temporal interfaces,

ARTICLE

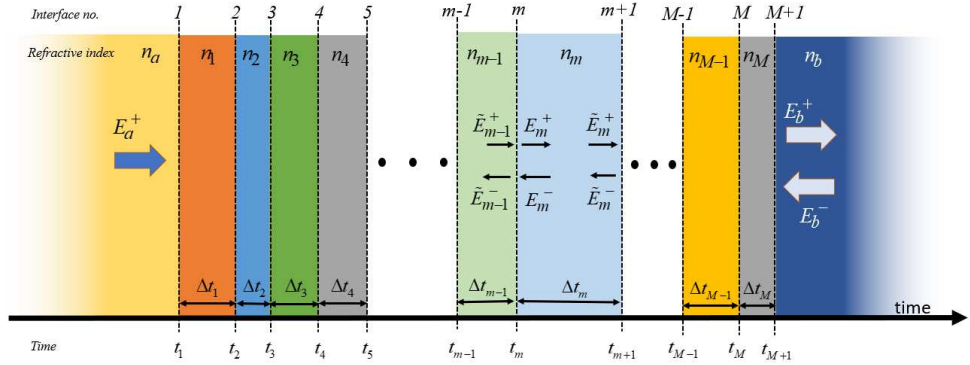

FIG. 1: General schematic representation of a temporal multilayer structure consisting of a cascade of $m$ temporal slabs between two temporally semi-infinite media with refractive indices $n_a$ and $n_b$, respectively.

showing that the amplitude of "reflected" and "refracted" waves across a temporal interface are proportional to the jump of refractive indices, formulating the temporal version of Fresnel equations. In this case, since the abrupt change of the refractive index involves the entire space, the "reflected" and "refracted" wave propagate in the same final medium.

A single interface does not provide enough degrees of freedom to tailor the overall scattering response, since it can generate only two waves, one backward- and one forward-propagating, at a different frequency with respect to the incident one. In [28], we proposed a double time-interface, realized by switching twice the properties of a uniform homogeneous medium in the whole space. It has been demonstrated that a double interface behaves as the temporal version of a dielectric slab between two semi-infinite media, i.e., a temporal slab, which allows designing both a perfect matching temporal slab, as proposed in [29], and a temporal Fabry-Perot cavity. The different response is achieved by simply selecting the proper effective application time of the slab and the refractive index. The study in [28], therefore, presents the design strategy to implement only first-order transfer functions [24].

In this Letter, we further extend the opportunities to tailor the frequency response of the scattered fields with temporal discontinuities by considering a generalized *temporal multilayer structure* (Fig. 1), where the cascade of different media is defined over time. The aim of this letter is to provide a rigorous analytical formulation and some design guidelines for special configurations to engineer the response of the temporal multilayers to tailor the desired frequency response, in analogy with what it can be achieved in conventional spatial multilayer structures to realize passband and stopband filters, matching networks, and multi-mode resonant cavities. To achieve this goal, we present a theoretical model and derive closed-form expressions of the transfer functions of such a temporal multilayer in terms of the overall scattering coefficients. The forward- and backward-wave scattering coefficients, relating the amplitudes of the corresponding waves to the one of the original forward propagating waves, are analytically derived as a function of frequency. We show that the response of the temporal multilayer can be controlled through the temporal sequence of refractive indeces and the application times of the structure, which act analogously to the electrical thickness of conventional spatial multilayer structure, allowing the realization of higher-order transfer functions for the electromagnetic response (Ch.5-6 of Ref. [24] for the definition of higher-order scattering structures). The analysis is conducted assuming an initial excitation, before the time switching, with harmonic time dependence $\exp[i\omega t]$, and arbitrary polarization.

Let us consider the multilayered temporal structure shown in Fig. 1, consisting of a cascade of $M$ different temporal slabs, between two semi-infinite media with refractive index $n_a, n_b$, respectively. Each temporal slab is characterized by the refractive index $n_m = \sqrt{\mu_m \varepsilon_m}$ and the application time $\Delta t_m$ (with $m = 1,2,3...M$), which is the time interval during which the $m$th medium is present in the whole volume where the wave is propagating. As demonstrated in [28] for a cascade of only two temporal interfaces, the overall scattering response is given by the interfering contributions in reflection and transmission from each interface, delayed in time by the corresponding application time. The analysis of the proposed structure can be, thus, decomposed in two separate stages: *i)* the first is devoted to describing the relationship between the fields before and after each temporal interface, i.e., the temporal matching conditions; *ii)* the second to properly delaying all the wave contributions present at a certain instant of time between two consecutive temporal interfaces, i.e., the delaying conditions introduced by each temporal slab, leading to the derivation of the corresponding transfer matrix [30].

*Temporal matching matrices* – Let us consider the $m$th interface of the temporal multilayer structure reported in Fig. 1. It is in common to the two media with refractive indices $n_{m-1}$, $n_m$. The temporal boundary conditions require that the total fields $D$ xxx, and $B$ be continuous across the two side of the interface [25]:

$$D = \tilde{D}$$
$$B = \tilde{B} \quad , \quad (1)$$

where the tilde symbol (~) over the fields identifies the quantities just before the interface (see Fig. 1). In terms of only the forward and backward D-field, boundary condition equations in (1) can be rewritten respectively as:

$$D_m^+ + D_m^- = \tilde{D}_{m-1}^+ + \tilde{D}_{m-1}^-$$
$$\eta_m(D_m^+ - D_m^-) = \eta_{m-1}(\tilde{D}_{m-1}^+ - \tilde{D}_{m-1}^-) \quad . \quad (2)$$

where $\eta_x = \sqrt{\mu_x/\varepsilon_x}$ is the wave impedance in the corresponding media ( $x = m, m-1$ ). Equation (2) may be written in a matrix form relating the fields *after* the temporal interface, $D^\pm$, to the ones *before* it, $\tilde{D}^\pm$:

$$\begin{bmatrix} D_m^+ \\ D_m^- \end{bmatrix} = \frac{1}{2}\begin{bmatrix} 1+\frac{\eta_{m-1}}{\eta_m} & 1-\frac{\eta_{m-1}}{\eta_m} \\ 1-\frac{\eta_{m-1}}{\eta_m} & 1+\frac{\eta_{m-1}}{\eta_m} \end{bmatrix}\begin{bmatrix} \tilde{D}_{m-1}^+ \\ \tilde{D}_{m-1}^- \end{bmatrix}. \quad (3)$$

Since all the materials are assumed to be linear, homogeneous, and isotropic, eq. (3) can be easily written also in terms of the propagating electric fields:

$$\begin{bmatrix} E_m^+ \\ E_m^- \end{bmatrix} = \begin{bmatrix} \tau_m & \rho_m \\ \rho_m & \tau_m \end{bmatrix}\begin{bmatrix} \tilde{E}_{m-1}^+ \\ \tilde{E}_{m-1}^- \end{bmatrix} \quad (4)$$

where $\tau_m, \rho_m$ are the temporal Morgenthaler transmission and reflection coefficients at the $m$th interface [25,26]:

$$\tau_m = \frac{1}{2}\left[\frac{\varepsilon_{m-1}}{\varepsilon_m} + \sqrt{\frac{\varepsilon_{m-1}\mu_{m-1}}{\varepsilon_m\mu_m}}\right]$$
$$\rho_m = \frac{1}{2}\left[\frac{\varepsilon_{m-1}}{\varepsilon_m} - \sqrt{\frac{\varepsilon_{m-1}\mu_{m-1}}{\varepsilon_m\mu_m}}\right], \quad (5)$$

which can be written also in terms of refractive index $n$ and intrinsic impedance $\eta$ of the adjacent $(m-1)$-th and $m$-th media as

$$\tau_m = \frac{1}{2}\left[\frac{n_{m-1}}{n_m}\frac{\eta_m + \eta_{m-1}}{\eta_{m-1}}\right]$$
$$\rho_m = \frac{1}{2}\left[\frac{n_{m-1}}{n_m}\frac{\eta_m - \eta_{m-1}}{\eta_{m-1}}\right] \quad (6)$$

Equation (4) relates the forward and backward propagating electric fields on one "side" of the temporal interface to those on the other "side", expressing the relationship in terms of a 2x2 temporal matching matrix MM at the $m$th interface:

$$MM_m = \begin{bmatrix} \tau_m & \rho_m \\ \rho_m & \tau_m \end{bmatrix}. \quad (7)$$

Once the forward and backward waves have passed across the $m$th interface, they propagate in the $m$th medium for the time period $\Delta t_m$, *i.e.*, the application time of the $m$th temporal slab.

*Delay matrices* – Similar to the phase delay due to propagation within the dielectric slabs, a temporal slab introduces a phase delay as well, but, in this case, it is related to the temporal phase of the propagating waves. The delay matrix describes the relationship between the electric fields just after an $m$th temporal interface and the ones just before the $(m+1)$th interface after the application time of the $m$th temporal slab.

Let us consider the electric fields at the temporal locations $t_m$ and $t_{m+1}$, separated by $\Delta t_m = t_{m+1} - t_m$. The forward fields at these two temporal locations can be written as:

$$E_m^+ = E_{m0}^+$$
$$\tilde{E}_m^+ = E_{m0}^+ e^{+i\xi_m\omega_0\Delta t_m} \quad , \quad (8)$$

where the $E_{m0}^+$ is the complex amplitude of the forward propagating wave just after the $m$th interface, $\omega_0$ is the original frequency in medium $n_a$, and $\xi_m = n_a/n_m$ is the time dilation factor introduced by the $m$th medium caused by changing of the wave velocity [25,28]. Exploiting the analogy with a spatial slab, we notice that the $m$th temporal slab introduces a phase delay similar to the one introduced by a spatial slab, where, however, the electrical thickness $d/\lambda$ is replaced by its effective temporal thickness $\xi_m\Delta t_m/T$, with $T = 2\pi/\omega_0$. In (8) the propagating term $\exp[-i\underline{k}_0 \cdot \underline{r}]$ is suppressed, since the temporal medium discontinuities maintain the wave momentum unaltered.

Similarly, $E_m^- = E_{m0}^-$ and $\tilde{E}_m^- = E_{m0}^- e^{-i\xi_m\omega_0\Delta t_m}$ for the backward propagating waves, leading to:

$$\begin{bmatrix} \tilde{E}_m^+ \\ \tilde{E}_m^- \end{bmatrix} = \begin{bmatrix} e^{+i\xi_m\omega_0\Delta t_m} & 0 \\ 0 & e^{-i\xi_m\omega_0\Delta t_m} \end{bmatrix}\begin{bmatrix} E_m^+ \\ E_m^- \end{bmatrix}, \quad (9)$$

where the anti-diagonal elements are zero due to the orthogonality of the propagating fields within the medium. Equation (9) relates the propagating electric fields in one temporal slab at two different time locations, thanks to the 2x2 temporal delay matrix:

$$DM_m = \begin{bmatrix} e^{+i\xi_m\omega_0\Delta t_m} & 0 \\ 0 & e^{-i\xi_m\omega_0\Delta t_m} \end{bmatrix}. \quad (10)$$

*Transfer matrix* – Based on the temporal matching and delay matrices, we can relate the incident field $E_a^+$ in the initial medium to the forward and backward propagating fields $E_b^+, E_b^-$ in the final one through the temporal transfer matrix as

$$\begin{bmatrix} E_b^+ \\ E_b^- \end{bmatrix} = \left(MM_1 \prod_{m=1}^{M} DM_m MM_{m+1}\right)\begin{bmatrix} E_a^+ \\ 0 \end{bmatrix} = TM_M \begin{bmatrix} E_a^+ \\ 0 \end{bmatrix}. \quad (11)$$

Regardless of the complexity of the temporal multilayered structure in terms of the number of slabs, the transfer matrix $TM_M$ is always composed of 4 terms (2x2) that represent the relative amplitude (at the original frequency $\omega_0$) of the final waves propagating in the medium $n_b$ at the new temporal frequency $\omega_b = \xi_{ba}\omega_0$ (see ref. [28,31]). Thanks to (11), we can derive the transfer matrix for a single temporal slab and compare the results with the one derived in [28]. The transfer matrix consists of only two matching matrices and one propagation matrix combined as

ARTICLE

$$TM_1 = \begin{bmatrix} \tau_1 & \rho_1 \\ \rho_1 & \tau_1 \end{bmatrix} \begin{bmatrix} e^{+i\xi\omega_0 \Delta t} & 0 \\ 0 & e^{-i\xi\omega_0 \Delta t} \end{bmatrix} \begin{bmatrix} \tau_2 & \rho_2 \\ \rho_2 & \tau_2 \end{bmatrix}. \quad (12)$$

Equation (12) returns the following complete matrix

$$TM_1 = e^{+i\xi\omega_0 \Delta t} \begin{bmatrix} \tau_1\tau_2 + \rho_1\rho_2 e^{-i2\xi\omega_0 \Delta t} & \tau_1\rho_2 + \rho_1\tau_2 e^{-i2\xi\omega_0 \Delta t} \\ \rho_1\tau_2 + \tau_1\rho_2 e^{-i2\xi\omega_0 \Delta t} & \rho_1\rho_2 + \tau_1\tau_2 e^{-i2\xi\omega_0 \Delta t} \end{bmatrix}$$

where it is easy to identify the expressions reported in Eqs. (5)-(6) in [28].

Although the application times $\Delta t_m$ may be arbitrarily selected, the analytical closed-form design equations can be derived only if the phase terms in the total scattering coefficients within the transfer matrix (11) have the form $\exp[\pm n\delta]$, where $n$ is a positive integer and $\delta$ is the phase delay introduced by each slab. This kind of temporal structures can be named as *equal travel-distance multilayered structures*, and it is characterized by the common propagation distance $L_s$ as follows:

$$v_m \Delta t_m = L_s \quad (m=1,2,3,...M) \quad (13)$$

where $v_m = c/n_m$ is the phase velocity of the wave contributions in the $m$th medium. From the delay matrix for the $m$th temporal slab in (10), we can identify the phase delay introduced by the application time

$$\delta = \omega \xi_m \Delta t_m = k v_m n_m \xi_m \Delta t_m = k n_a L_s, \quad (14)$$

where the definitions $\omega = kc$, $c = v_m n_m$, $\xi_m = n_a/n_m$ and Eq. (13) have been used. Being the wave momentum $k = 2\pi/\lambda$ conserved passing through the temporal metamaterial, we can rewrite (14) as

$$\delta = 2\pi \frac{L_s}{\lambda}. \quad (15)$$

It is clear from (15) that the characteristic distance $L_s$ of the equally travel-distance multilayered structure plays a special role. If the characteristic distance is chosen to be $\lambda/2$ and $\lambda/4$, the phase delay introduced by each temporal slab is $\pi$ and $\pi/2$ respectively, at the frequency for which the operating wavelength is $\lambda$, regardless the refractive indices of the temporal slabs. Such special conditions are achieved by designing a temporal multilayered structure characterized by the application times

$$\Delta t_m = \frac{1}{2\pi} \frac{n_m}{n_a} \delta T_0. \quad (16)$$

When $\delta = \pi$, the entire temporal multilayered structure is transparent to the propagating wave due to the destructive interference of wave contributions in both forward and backward directions within it, regardless of the values of the refractive indeces of the media constituting the temporal multilayered structure, and the transfer matrix is reduced to the matching matrix between the initial and final media

$$TM_\pi = \begin{bmatrix} \tau_{ab} & \rho_{ab} \\ \rho_{ab} & \tau_{ab} \end{bmatrix}, \quad (17)$$

where $\tau_{ab}$ and $\rho_{ab}$ can be evaluated using (5) or (6), substituting the subscripts with $m-1 \to a$, $m \to b$.

On the contrary, when $\delta = \pi/2$, the contributions add in-phase within the temporal slabs, leading to a gain effect at the design frequency. The final forward and backward waves will be amplified with respect to the incident one, as shown in the following. It is important to stress that, as opposed to spatial interfaces, at time interfaces energy is not generally conserved, since an index change of a polarized medium generally requires or dissipates energy.

*High-order transfer function engineering*– Cascading several temporal slabs is our strategy to increase the available degrees of freedom and tailor the frequency response of the temporal multilayered structure. Like in conventional spatial multilayered structures, the different temporal slabs act as resonators at different frequencies, and their sequential cascade can be used to tailor the overall frequency response of the system. The proper selection of the design parameters is fundamental for the synthesis of the desired frequency response, especially in the case of complex nonconventional transfer functions.

To demonstrate the possibility of designing high-order transfer functions, we consider here three relevant examples belonging to the four classes of multilayered devices that can be designed by exploiting the effective temporal thicknesses of the slabs. We consider here non-magnetic multilayered structures ($\mu_a = \mu_b = \mu_m = 1$), showing the potential to tailor the frequency scattering response of the structure:

1) a temporal multilayer consisting of 4 temporal slabs with arbitrary refractive indices and application times between two different temporally semi-infinite media.
2) an equally travel-distance multilayered structure with $L_s = \lambda/2$ ($\delta = \pi$) with arbitrary refractive indices.
3) an equally travel-distance multilayered structure with $L_s = \lambda/2$ ($\delta = \pi$) with periodic cascade of only two media with high and low refractive indices $n_H, n_L$, respectively.
4) an equally travel-distance multilayered structure with $L_s = \lambda/4$ ($\delta = \pi/2$), but same material configuration of Example 3.

The electromagnetic response over a wide frequency bands is predicted by the propagation and scattering model reported in this Letter in terms of final forward and backward scattering coefficients as a function of normalized frequency. Our analytical results are also compared with full-wave FDTD numerical simulations. The FDTD code has been developed ensuring stability and convergence of the simulations, even when the abrupt switching of the material properties occurs. The propagation in temporal metamaterials is characterized by a constant wavevector, but a continuous change of the temporal frequency, which may assume instantaneously much higher values than the initial one. Therefore, it is fundamental to start the discretization of propagation domain from the temporal dimension imposing a time step $dt$ smaller enough to capture the highest frequency generated by the switching. Once the temporal grid is defined, the spatial discretization can be performed according to the Courant stability criterion, by imposing a $dz > dt\, c/n_{\min}$. Increasing the number of nodes per time-period increases the mapping accuracy of the time location of the $m$th temporal interface within the discretized grid of the FDTD simulation, at the expense of longer computation time, as expected in any numerical simulation. In the following, the frequency responses of the four examples are reported.

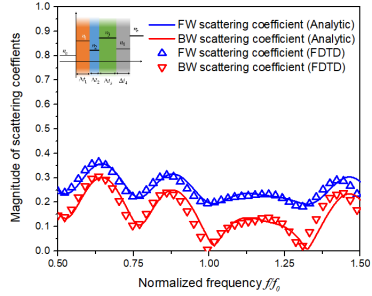

FIG. 2: Example 1 – Comparison between analytical and numerical scattering coefficients as a function of frequency of the arbitrary temporal multi-layered structure shown in the inset. The parameters are reported in Table I.

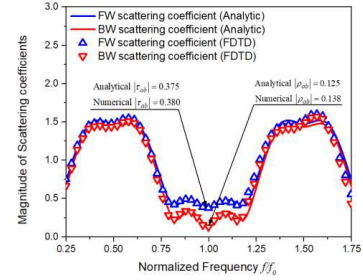

FIG. 3: Example 2 – Comparison between analytical and numerical scattering coefficients as a function of frequency of an equally travel-distance multi-layered structure with $L_s = \lambda/2$ $(\delta = \pi)$. Refractive indices are: $n_a = 1$, $n_1 = 2$, $n_2 = 7$, $n_3 = 1.5$, $n_4 = 5$, $n_b = 2$.

Example 1 - Multi-layered arbitrary device. The first multilayered device consists of a cascade of four temporal slabs whose refractive indices are reported in Table I. The initial medium where the incident wave is propagating is vacuum ($n_a = 1$), whereas the final medium has a refractive index $n_b = 3$. The application time of each temporal slabs is defined as a function of an arbitrary time $T_0$, which determines the central frequency $f_0 = 1/T_0$ of the frequency range over which the forward and backward scattering coefficients are analytically and numerically computed.

TABLE I. Example I: Arbitrary multi-layered structure with 4 temporal slabs, whose scattering parameters are reported in Fig. 2.

| Medium identifier | Refractive Index $n = \sqrt{\varepsilon_r}$ ($\mu_r = 1$) | Application Time $(\Delta t_m/T_0)$ |
|---|---|---|
| a | 1.0 | ---- |
| 1 | 2.5 | 0.5 |
| 2 | 4.4 | 1 |
| 3 | 2.0 | 3.5 |
| 4 | 1.6 | 0.25 |
| b | 3.0 | ---- |

In Fig. 2, we report the comparison between the analytical and numerical results of this multilayered structure. The results agree very well over the entire frequency range. The scattering coefficients are below unity in this specific example, implying that the wave has lost energy during the switching cascade. For different switching times, the scattering coefficients may acquire values greater than unity, revealing a gain effect induced on the propagating signal by the medium. This will be clearly shown in the next examples.

Example 2 - Transparent multi-layered device. We now consider an equally travel-distance multilayered device with $L_s = \lambda/2$ $(\delta = \pi)$, whose temporal slabs have the following arbitrary refractive indices: $n_a = 1$, $n_1 = 2$, $n_2 = 7$, $n_3 = 1.5$, $n_4 = 5$, $n_b = 2$. In Fig. 3 we report the comparison between analytical and numerical scattering coefficients as a function of frequency of such an equally travel-distance multi-layered structure in the frequency range $[0.5f_0, 1.5f_0]$, where $f_0$ is the frequency at which the temporal multi-layered structure is designed to have $\delta = \pi$. At $f = f_0$, due to the destructive interference between all the waves generated by the interfaces of the multilayered structure, the total forward and backward scattering coefficients are defined only by the discontinuity introduced by the different initial and final medium. In Fig. 3, in addition to the scattering parameters as a function of frequency, we report also the comparison between the values of the forward and backward scattering coefficients at $f = f_0$ computed using eq. (17) and the numerical FDTD simulator. They agree very well, demonstrating the transparency of the structure at $f_0$ and over a moderately broad bandwidth around it.

Example 3 - Periodical multi-layered device. The previous device can be further simplified by considering an equally travel-distance multilayered structure with $L_s = \lambda/2$ $(\delta = \pi)$, whose temporal slabs are a periodic cascade of only two media with high and low refractive indices $n_1 > n_2$. This configuration is particular of interest for its periodicity that affects the overall scattering response, creating a time crystal. In Fig. 4, we report the comparison between analytical and numerical scattering coefficients as a function of frequency. As in Example 2, again with $L_s = \lambda/2$ $(\delta = \pi)$ at $f = f_0$, the temporal multi-layered structure does not introduce any additional scattering contribution at this frequency. However, here we consider the special configuration in which the initial and final media have the same refractive indeces $n_a = n_b$ and the same intrinsic impedances $\eta_a = \eta_b$, and $\mu = 1$ by initial assumption. Therefore, the forward and backward scattering coefficients approach unity and zero, respectively, as predicted by Eq. (17). Moreover, due to the

ARTICLE

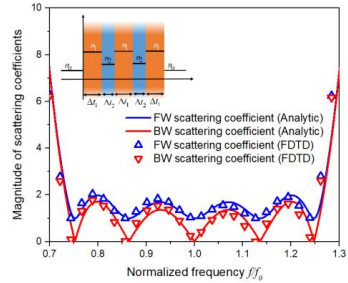

FIG. 4: Example 3 – Comparison between analytical and numerical scattering coefficients as a function of frequency of an equally travel-distance multi-layered structure with $L_s = \lambda/2$ ($\delta = \pi$), consisting of five temporal slabs (inset figure) whose refractive indices are: $n_a = n_b = 1$, $n_1 = 6$, $n_2 = 1.5$.

periodicity of the structure, the vanishing of all scattered waves at the interfaces takes place at a discrete number of frequencies around the design frequency $f_0$, strictly related to the number of temporal slabs constituting it. This feature is found also in conventional spatial multi-layered structures, showing the connection between the designs of temporal and spatial multilayer devices. This demonstration allows to further extend the design of temporal devices by applying the well-established design strategies used for synthetizing frequency responses in filter theory [24,32]. Finally, we discuss the response outside the band of interest. In this specific example, the scattering coefficients grow well above unity within two consecutive zeros of the reflection coefficient (or unity of the transmission coefficient), as a testimony of the possibility of pumping energy in the wave

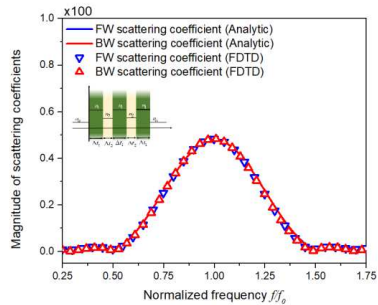

FIG. 5: Example 4 – Comparison between analytical and numerical scattering coefficients as a function of frequency of an equally travel-distance multi-layered structure with $L_s = \lambda/4$ ($\delta = \pi/2$), consisting of five temporal slabs (inset figure) whose refractive indices are: $n_a = n_b = 1$, $n_1 = 6$, $n_2 = 1.5$.

through switching, analysed in more detail in the next example.

Example 4 - Amplifying multi-layered device. The last device considered here an equally travel-distance multilayered structure with $L_s = \lambda/4$ ($\delta = \pi/2$) and with same material configuration as in Example 3, that is a cascade of high and low refractive indices. The Bragg grating, or dielectric mirror, is one of the basic devices in optics and electromagnetics, used in substitution of ordinary metallic mirrors for reflecting optical and electromagnetic signals with extremely low losses. However, in the case of temporal multilayered structure with $L_s = \lambda/4$ ($\delta = \pi/2$), the wave contributions both in forward and backward directions generated at each temporal discontinuity do not interfere as in conventional Bragg grating, but they are always in-phase and gain energy from the system at each temporal discontinuity, leading to the generation of two amplified copies of the incident signal propagating in both forward and backward directions. This is clearly shown in Fig. 5, where we report the amplitude of the scattering parameters (scale x100), and the comparison of the results obtained though the analytical model and numerical FDTD simulation, respectively.

To conclude, in this Letter we have demonstrated the possibility to engineer the frequency response of the scattered fields from a temporal multilayer structure, consisting of a cascade of different media over time, instead of over space. In analogy with conventional spatial multilayer structures, we have demonstrated that the frequency response can be properly engineered by controlling the refractive indices and application times of the temporal slabs, enabling also special scattering responses, such as "transparent" and "pumping" temporal multilayered structures. Moreover, thanks to the theoretical model proposed in this letter, we have shown that the design strategies to tailor the frequency response of our temporal multilayers can be derived by well-established design strategies used for synthetizing frequency responses in filter theory, allowing to synthetize complex transfer functions. Special high-order transfer functions, such as maximally-flat and Chebyshev ones used in filter design theory, may be applied to temporal multi-layered structures following similar principles.

### DATA AVAILABILITY

The data that support the findings of this study are available within this article. Numerical FDTD codes developed in MATLAB can be provided at the email of the corresponding author upon a reasonable request.

### ACKNOWLEDGEMENTS

D.R., F.B. acknowledge the financial support of the Italian Ministry of Education, University and Research as a PRIN 2017 project (research contract MANTLES - protocol number 2017BHFZKH). A.A. acknowledges the financial support of the Air Force Office of Scientific Research and Simons Foundation.

xxx,